# SUPER MULTI-INSTANTONS IN CONFORMAL CHIRAL SUPERSPACE

W. Siegel[1]

*Institute for Theoretical Physics*
*State University of New York, Stony Brook, NY 11794-3840*

## ABSTRACT

We reformulate self-dual supersymmetric theories directly in conformal chiral superspace, where superconformal invariance is manifest. The superspace can be interpreted as the generalization of the usual Atiyah-Drinfel'd-Hitchin-Manin twistors (the quaternionic projective line), the real projective light-cone in six dimensions, or harmonic superspace, but can be reduced immediately to four-dimensional chiral superspace. As an example, we give the 't Hooft and ADHM multi-instanton constructions for self-dual super Yang-Mills theory. In both cases, all the parameters are represented as a single, irreducible, constant tensor.

---

[1]　Internet address: siegel@insti.physics.sunysb.edu.

# 1. INTRODUCTION

Attention has returned recently to self-dual Yang-Mills theory for various reasons, mostly in relation to the properties of lower- or higher-dimensional theories [1]. In this paper we will be concerned mainly with its original use for studying features of Yang-Mills theory and its supersymmetric generalizations in four dimensions, and in particular conformal invariance. Our method is a new supertwistor formulation of self-dual theories. The twistors are essentially the same kind as those used in the general multi-instanton construction, but we elevate them to the status of fundamental off-shell coordinates on which all self-dual theories are defined. We apply this approach to supersymmetric multi-instanton solutions, for all numbers (N=1 to 4) of supersymmetries. Our expressions are simpler than previous results, being *identical* in form to the bosonic ones, with just a change in the range of the indices.

We begin by discussing alternative formalisms for conformal theories: the original (super)twistors, which are inherently on-shell and tied to Minkowski space, and the projective light-cone, which can describe conformal theories in any dimension. We then introduce self-dual twistors, which are off-shell and particularly suited for self-dual theories. They appear as the square root of the projective light-cone coordinates. We give the general formulation of self-dual particle and free field theories in this twistor space, and discuss interactions (especially Yang-Mills theory). Although previously Minkowski and self-dual twistors have been treated as almost identical, their differences are crucial in understanding this formulation. In particular, self-dual twistors reduce to the usual four-dimensional $x$ coordinates upon a simple gauge choice. As an example, we write a new action for the massless particle that (1) is manifestly conformal, (2) requires self-dual (not Minkowski) twistors, (3) uses no square roots or Lagrange multipliers (only the twistors, no world-line metric, etc.), and (4) gives the usual propagator upon four-dimensional quantization.

The space is also a slight generalization of harmonic superspace, and harmonic methods can be applied to self-dual twistors to give a more natural description of SU(2) instantons than either the usual four-dimensional space or the projective light-cone: The self-dual twistor coordinates, being a spinor representation of SO(5,1) or SO(3,3), automatically carry both a four-valued conformal index of SU*(4) or SL(4) (the conformal group) and a two-valued index of SU(2) or SL(2) (an internal group). This two-valued index, although arising as a result of representing the conformal group, can be tied to the SU(2) Yang-Mills group in the same way as is usually done for half of the Lorentz group SO(4)=SU(2)⊗SU(2) (or SL(2) Yang-Mills for SO(2,2)=SL(2)⊗SL(2)). Thus both conformal and Yang-Mills SU(2) invariances are manifest.

Next, we review the construction of arbitrary multi-instantons for arbitrary Yang-Mills groups in a way that emphasizes the role of self-dual twistors for both SO(4) and SO(2,2) spacetimes. We then consider the simplest formalism for describing self-dual supersymmeric theories in four-dimensional space: In chiral superspace the trivial half of the supersymmetries are treated as part of the manifest SL(N|2)⊗SL(2) "super-Lorentz" symmetry, while the other half are treated as part of the translations in the (N+2)×2 supercoordinates of a torsion-free superspace. Not only is this superspace simpler than the usual superspace, but it is also much more similar to ordinary space. The generalization to self-dual supertwistors is immediate: Simply increase the range of the super-index, since superconformal symmetry is SL(N|4)⊗SL(2). This conformal chiral superspace formalism then allows the automatic supersymmetrization of multi-instanton constructions.

## 2. MINKOWSKI TWISTORS

Originally [2] twistors were introduced to represent the conformal group in 3+1 dimensions, SO(4,2). Essentially, you start with a complex four-component bosonic spinor, the defining representation of SU(2,2) (=SO(4,2)), and define its complex conjugate to also be its canonical conjugate:

$$[z_a, \bar{z}^b] = \delta_a^b$$

This gives the conformal generators via the usual oscillator construction of SU(m,n):

$$J_a{}^b = \bar{z}^b z_a - \text{trace}$$

(This construction is closely related to that of the $\gamma$-matrix representation of SO(m,n). Similar constructions apply to 2+1 and 5+1 dimensions [3,4].)

Since the translations are part of conformal transformations, the free massless equation of motion $p^2 = 0$, and all equations related to it by conformal boosts, can be simply represented in general as [5]

$$J_{[a}{}^{[c} J_{b]}{}^{d]} - \text{trace} = 0$$

([ ] stands for antisymmetrization, ( ) for symmetrization.) This equation is satisfied identically by the twistor representation. (This is related to the fact that spinors are conformal.) In particular, if we reduce to four-dimensional notation (SU(2,2)→SL(2,C)) by writing $z_a = (z_\mu, z_{\dot\mu})$, $\bar z^a = (\bar z^\mu, \bar z^{\dot\mu})$, then $p_{\mu\dot\mu} = z_\mu \bar z_{\dot\mu}$ automatically satisifies $p^2 \equiv \frac{1}{2} C^{\mu\nu} \bar C^{\dot\mu\dot\nu} p_{\mu\dot\mu} p_{\nu\dot\nu} = 0$ because of the antisymmetry of the SL(2,C) metric $C$ ($=\sigma_2$) and the commutativity of the bosonic variables $z$.

The twistor represents four real variables (and their canonical conjugates): the three independent components of the on-shell momentum and the single component of the helicity, whose operator

$$h = \bar z^a z_a$$

is the usual chiral U(1) symmetry of spinors for SO(3,1) and SO(4,2).

The generalization to conformal supersymmetry is straightforward [6,4]: Just grade the group. The superconformal group in 3+1 dimensions is SU(N|2,2) for N supersymmetries, so we replace the bosonic index $a$ with the super index $A = (a, a')$, introducing N fermionic oscillators $z_{a'}$.

The Penrose transform can be used to express free on-shell fields in terms of fields on twistor space: By using the usual Fourier representation of such fields, and "solving" the $\delta(p^2)$ factor by replacing $p$ with its twistor expression, a scalar field can be expressed as

$$\phi(x) = \int d^2 z_\mu d^2 \bar z_{\dot\mu} \; exp(ix^{\mu\dot\mu} z_\mu \bar z_{\dot\mu}) \hat\phi(z_\mu, \bar z_{\dot\mu})$$

$$= \int d^2 z_\mu \; \tilde\phi(z_\mu, x^\mu{}_{\dot\mu} z_\mu)$$

$$= \int d^4 z_a \; \delta(z_{\dot\mu} - x^\mu{}_{\dot\mu} z_\mu) \tilde\phi(z_a)$$

This representation generalizes to describe arbitrary free field strengths by including arbitrary numbers of $z_\mu$ and $\bar z_{\dot\mu}$ factors in the integrand (since $p^{\mu\dot\mu} z_\mu = p^{\mu\dot\mu} \bar z_{\dot\mu} = 0$). However, since $p_{\mu\dot\mu} = z_\mu \bar z_{\dot\mu}$ implies the energy $\sum_\mu |z_\mu|^2$ is positive definite, $\phi$ describes only positive energy, and $\bar\phi$ negative energy. Similarly, first-quantized path-integral representations of this transform, from classical mechanics lagrangians [7] such as $\dot x^{\mu\dot\mu} z_\mu \bar z_{\dot\mu}$, give propagators $\Theta(p_0)\delta(p^2)$ instead of $1/p^2$. This is related to the fact that such lagrangians are not positive definite: They follow from the usual lagrangian $\frac{1}{2} g \dot x^2$ by treating $g$ as an ordinary Lagrange multiplier, whereas the usual $1/p^2$ propagator follows only if $g$ (and thus the action) is restricted to be positive definite.

## 3. PROJECTIVE LIGHT-CONE

An alternative to twistors for manifesting conformal invariance is to work on the projective light-cone in two extra dimensions (one space, one time) [8]. The basic idea is to start with a vector coordinate for the group, constrain its square to vanish, and allow a scale invariance to gauge away the other extra component. For example, if we write the particle Lagrangian [9] (with action $S = \int d\tau \, L$)

$$L = \tfrac{1}{2}\dot{y}^2 - \tfrac{1}{2}\lambda y^2$$

and vary the Lagrange multiplier $\lambda$ to enforce the constraint $y^2 = 0$ through the solution

$$y = ew, \quad w = (1, -\tfrac{1}{2}x^2, x) \quad \Rightarrow \quad w^2 = w \cdot \dot{w} = 0$$

(where the first two components of $w$ are the $\pm$ components in terms of the two extra dimensions) in terms of the (in this case) four-vector $x$, then we find

$$\tfrac{1}{2}\dot{y}^2 = \tfrac{1}{2}(e\dot{w} + \dot{e}w)^2 = \tfrac{1}{2}e^2\dot{w}^2 = \tfrac{1}{2}e^2\dot{x}^2$$

which is the usual lagrangian in terms of the ordinary coordinates $x$ and the world-line einbein $e$. The projective light-cone thus "unifies" $x$ and $e$ into $y$. Unlike the twistor actions mentioned above, this action is positive definite. It can be world-line supersymmetrized straightforwardly [9] to describe arbitrary conformal representations [10]. Since this action implies an SL(2) algebra of constraints $y^2$, $y \cdot \partial/\partial y$ (scale invariance), and $(\partial/\partial y)^2$ (Klein-Gordon equation), $y$ and $\partial/\partial y$ are treated on an equal basis (in the free theory). In particular, the extra two dimensions are eliminated in a way analogous to the way that the Klein-Gordon equation (or field equations in general) "eliminate" time as an independent dimension. As a result, the use of such formalisms can simplify theories in the same way that four-component notation (and the use of Lagrangians) introduces simplifications over three-component notation (and the use of Hamiltonians) in Lorentz invariant theories.

Similar methods can be applied to field theory [8,11]. In the case of Yang-Mills theory, the basic idea is to covariantize the conformal generators, which preserve the projective light-cone, since they commute with the SL(2) algebra of constraints. In particular, for Yang-Mills theory we have the covariant derivative

$$\nabla_{\mathcal{AB}} = y_{[\mathcal{A}}\nabla_{\mathcal{B}]}, \quad \nabla_{\mathcal{A}} = \partial_{\mathcal{A}} + A_{\mathcal{A}}$$

which is covariant under the usual gauge transformations as well as invariant under the independent gauge transformations

$$\delta A_{\mathcal{A}} = y_{\mathcal{A}} \hat{\Lambda}$$

In addition, for scale invariance to be field independent we require a constraint that resembles a Lorentz gauge condition:

$$y^{\mathcal{A}} \nabla_{\mathcal{A}} = y^{\mathcal{A}} \partial_{\mathcal{A}} \quad \Rightarrow \quad y^{\mathcal{A}} A_{\mathcal{A}} = 0$$

The extra gauge invariance and constraint reduce the six-component $A$ to the usual four-vector. The field strength $F_{\mathcal{ABC}}$ is defined by

$$[\nabla_{\mathcal{AB}}, \nabla_{\mathcal{CD}}] = -y_{[\mathcal{A}} F_{\mathcal{B}]\mathcal{CD}} \quad \Rightarrow \quad F_{\mathcal{ABC}} = \tfrac{1}{2} y_{[\mathcal{A}} F_{\mathcal{BC}]}, \quad [\nabla_{\mathcal{A}}, \nabla_{\mathcal{B}}] = F_{\mathcal{AB}}$$

Only $\nabla_{\mathcal{AB}}$ and $F_{\mathcal{ABC}}$ (not $\nabla_{\mathcal{A}}$ and $F_{\mathcal{AB}}$) are covariant under both gauge transformations. Furthermore, a self-duality condition can be imposed on the field strength (for appropriate spacetime signatures):

$$F_{\mathcal{ABC}} = \pm \tfrac{1}{6} \epsilon_{\mathcal{ABCDEF}} F^{\mathcal{DEF}}$$

which is equivalent to the usual four-dimensional condition.

More details of this approach will be described below, when we give a more thorough discussion of the related formalism of self-dual twistors.

## 4. SELF-DUAL TWISTORS

Self-duality for four-dimensional theories (along with the usual reality properties) requires an even number of time dimensions. The relevant conformal group is then either SO(5,1) (=SU*(4)) or SO(3,3) (=SL(4)). For these groups spinor indices cannot be raised or lowered, so the above twistor construction of the generators fails. However, one again can be led to twistors from a different analysis:

The use of the constraint $y^2 = 0$ for the projective light-cone is closely analogous to the way $p^2 = 0$ is treated in the usual light-cone formalism. This suggests as an alternative the use of twistors for $y$:

$$y^{ab} = z^{a\alpha} z^b{}_\alpha$$

where $y$ is antisymmetric in its two spinor indices (the six-component vector representation). However, because of reality properties of spinors for SO(n,6-n), this works only for the same cases as self-duality. Specifically, for SO(5,1) $z$ is a real spinor representation of SU*(4)⊗SU(2) (it is a pseudoreal representation of SU*(4) and of SU(2)), while for the simpler case of SO(3,3) it is a real spinor representation of SL(4)⊗SL(2) (a real representation of each). In both cases it has eight real components, but scale transformations together with SU(2) or SL(2) transformations on the $\alpha$ index eliminate half the components. This SU(2) replaces the U(1) of Minkowski twistors, but unlike the Minkowski case this symmetry has no physical significance. Also unlike the Minkowski case, where $z$ is canonically conjugate to its complex conjugate, here the $z$'s all commute, and so we must introduce the $z$-derivatives $\partial_{a\alpha}$. These four transformations (constraints) are represented by the operators

$$J_\beta{}^\alpha = z^{a\alpha}\partial_{a\beta}$$

which commute with the conformal generators

$$J_b{}^a = z^{a\alpha}\partial_{b\alpha} - \text{trace}$$

Furthermore, these twistors are off-shell, also unlike Minkowski twistors. Field equations can be written as generalizations of the usual four-dimensional field equations for conformal field theories. For example, the Klein-Gordon equation is

$$\partial_a{}^\alpha \partial_{b\alpha} = 0$$

By comparison with the above discussion of the projective light-cone, we can find the relation to the usual four-dimensional coordinates. Similarly to Minkowksi twistors, the reduction of indices is $a = (\mu, \mu')$, but $\mu'$ is an independent index (not complex conjugate), since SO(4)=SU(2)⊗SU(2) and SO(2,2)=SL(2)⊗SL(2) instead of SO(3,1)=SL(2,C).

$$z^a{}_\alpha = \lambda_\alpha{}^\nu(\delta^\mu_\nu, x_\nu{}^{\mu'})$$

gives the same expression for $y$ in terms of $x$ as above, with the identification $e = \lambda^2$. The equivalent relation $z^{\mu'}{}_\alpha = x_\nu{}^{\mu'} z^\nu{}_\alpha$ corresponds to the equation $z_{\dot\mu} = x^\mu{}_{\dot\mu} z_\mu$ for Minkowski twistors, but with the important difference that here it can be solved for $x$ in terms of the $z$'s, since $z^\nu{}_\alpha = \lambda_\alpha{}^\nu$ is an invertible 2×2 matrix ($e \neq 0$). If we change variables to $x$ and $\lambda$, the conformal generators take the usual nonlinear form with $\lambda$ appearing only as the (chiral) spin and scale weight operators $\lambda \partial/\partial \lambda$.

This choice of variables is directly related to the twistors used in the Atiyah-Drinfel'd-Hitchin-Manin (ADHM) construction of all multi-instanton solutions [12]. To translate into quaternion langauge, we identify any 2×2 matrix as a quaternion:

Then $z$ is a doublet of quaternions, while $x$ is their ratio, describing the quaternionic projective line. The quaternion $\lambda$ can be gauged arbitrarily by the symmetry SU(2)⊗GL(1) or GL(2) described above. (Actually, an element of the 2×2 matrix representation of SU(2)⊗GL(1) *is* a quaternion; it's determinant is the square of an SO(4) vector. In the GL(2) case we get a Wick rotation of a quaternion: The determinant of a real 2×2 matrix is the square of an SO(2,2) vector.) This differs from Minkowski twistors, where the gauge symmetry is only U(1)⊗GL(1) (a complex scale transformation), so the natural language is in terms of a *complex* projective space. While ADHM used quaternions as an incidental feature of their construction, and in a way that did not illustrate their distinction from Minkowski twistors, we will formulate general self-dual theories directly in this space, and make use of both their SU*(4) (SL(4)) and SU(2) (SL(2)) symmetries.

## 5. PARTICLES AND FREE FIELDS

A classical mechanics Lagrangian for spin zero can be obtained from that for the projective light-cone by just substituting $y = z^2$:

$$L = -\tfrac{1}{4}\epsilon_{abcd} z^{a\alpha} z^b{}_\alpha \dot{z}^{c\beta} \dot{z}^d{}_\beta$$

where we have used the SU*(4) or SL(4) form of the SO(5,1) or SO(3,3) metric, $y^2 = \tfrac{1}{4}\epsilon_{abcd} y^{ab} y^{cd}$. The action is invariant under local SU(2) or SL(2) transformations, as well as under reparametrizations $\delta z = \xi \dot{z} - \tfrac{1}{4}\dot{\xi} z$. In the gauge $\lambda_\alpha{}^\nu = e^{1/2}\delta_\alpha^\nu$ it gives the usual four-dimensional action (unlike Minkowski twistor actions). This new action for the particle is thus unique in that it is reparametrization invariant without the use of square roots or Lagrange multipliers, while still giving the usual propagator upon quantization.

An alternative approach to first-quantization that describes arbitrary spin is to write a set of constraints directly in operator form. The constraints can be expressed in terms of the spin generators (acting only on indices of fields or wave functions) of SU*(4)⊗GL(1) or GL(4). (This corresponds to the way nonrelativistic quantum mechanics is usually done, using spin operators in the Hamiltonian instead of expressing the spin in terms of classical variables.) One way to derive them is to start with the projective light-cone constraints [10] and use the relations

$$y^{ab} = z^{a\alpha} z^b{}_\alpha, \quad \partial_{a\alpha} = 2 z^b{}_\alpha \partial_{ab}$$

(The latter relation holds on all functions that depend on $z$ only through $y$.) We then find

$$y^{ab}M_b{}^c = 0 \quad \Rightarrow \quad z^{a\alpha}M_a{}^b = 0$$
$$\{y^{ab}, \partial_{ab}\} = 0 \quad \Rightarrow \quad \{z^{a\alpha}, \partial_{a\beta}\} = 0$$
$$\partial^{ab}M_b{}^c = 0 \quad (\partial_{[ab}M_{c]}{}^d = 0) \quad \Rightarrow \quad \partial_{[a\alpha}M_{b]}{}^c = 0$$
$$\partial^{ab}\partial_{ab} = 0 \quad (\partial_{[ab}\partial_{cd]} = 0) \quad \Rightarrow \quad \partial_a{}^\alpha \partial_{b\alpha} = 0$$

where $M_a{}^b$ are the SU*(4)⊗GL(1) or GL(4) generators. The former two equations are kinematic, and restrict the indices and coordinates, respectively, to the usual four-dimensional ones. The latter two are dynamic, and further restrict the indices and coordinates to the usual on-shell (light-cone) ones. There is also a constraint unaffected by the change of coordinates

$$M_{[a}{}^{[c}M_{b]}{}^{d]} - \text{trace} = 0$$

which restricts the representations to conformal ones. Except for scalars these representations are exactly the ones that have both self-dual and anti-self-dual parts. However, the $zM$ constraint then picks out just the self-dual representations.

All self-dual field strengths are totally symmetric in lower spinor indices: $F_{(a\ldots b)}$. (The anti-self-dual ones are totally symmetric in upper spinor indices.) The independent constraints on field strengths are then, for nonzero spin,

$$z^{a\alpha}F_{a\ldots b} = 0, \quad \partial_{[a\alpha}F_{b]\ldots c} = 0$$

These imply

$$z^{a\alpha}\partial_{a\beta}F_{b\ldots c} + \delta^\alpha_\beta F_{b\ldots c} = 0, \quad \partial_a{}^\alpha \partial_{b\alpha}F_{c\ldots d} = 0$$

which are also the independent constraints for spin zero.

## 6. INTERACTIONS

Generalization to interactions is straightforward. The main difference from the usual four-dimensional formalism is checking that the kinematic constraints are preserved by the interactions. For example, for a self-interacting scalar (massless $\phi^4$ theory)

$$\partial_a{}^\alpha \partial_{b\alpha}\phi = y_{ab}\tfrac{1}{6}\phi^3$$

The form of the potential was determined by the $z\partial$ constraint, which fixes the GL(1) weight of all field strengths.

For a gauge theory the constraints are modified by replacing the partial derivatives with covariant derivatives $\nabla_{a\alpha} = \partial_{a\alpha} + A_{a\alpha}$. To keep the kinematic constraints kinematic under this generalization, we require

$$z^{a\alpha}\nabla_{a\beta} = z^{a\alpha}\partial_{a\beta} \quad \Rightarrow \quad z^{a\alpha}A_{a\beta} = 0$$

However, this constraint is a special case of the other kinematic constraint, $z^{a\alpha}M_a{}^b = 0$. On the other hand, the coordinate restriction must be modified, since the gauge fields covariantize the $z$ coordinates, which differ from the projective light-cone gauge fields by a factor of $z$:

$$A = dy^{ab}A_{ab} = dz^{a\alpha}A_{a\alpha} \quad \Rightarrow \quad A_{a\alpha} = 2z^b{}_\alpha A_{ab}$$

$$J_a{}^b = 2y^{bc}\nabla_{ac} - \text{trace} = z^{b\alpha}\nabla_{a\alpha} - \text{trace} \quad \Rightarrow \quad 2y^{bc}A_{ac} = z^{b\alpha}A_{a\alpha}$$

The former equation gives $A_{a\alpha}$ explicitly in terms of $A_{ab}$, while the latter (which is actually a consequence of the former) gives the $\hat{\Lambda}$-gauge-independent part of $A_{ab}$ in terms of $A_{a\alpha}$. ($J_a{}^b$ is the $\nabla_{\mathcal{AB}}$ used earlier.)

The remaining kinematic constraint on the gauge field follows from the $zA$ constraint together with the corresponding constraint $zF$ on its field strength. We then obtain generalized kinematic constraints that can be applied to all fields, whether field strengths, gauge fields, or gauge parameters:

$$z^{a\alpha}M_a{}^b = 0, \quad J_\beta{}^\alpha \equiv z^{a\alpha}\partial_{a\beta} + M_\beta{}^\alpha = 0$$

where $M_\alpha{}^\beta$ are the "spin generators" for SU(2)⊗GL(1) or GL(2). In fact, the latter constraint is simply the restriction to singlets of the full group generators (orbital + spin). In particular, the gauge field satisfies

$$z^{a\alpha}\partial_{a\beta}A_{b\gamma} + \delta^\alpha_\gamma A_{b\beta} = 0$$

Finally, the dynamic equation for gauge theories is the equation of self-duality

$$[\nabla_{a\alpha}, \nabla_{b\beta}] = C_{\beta\alpha}F_{ab}$$

or in differential form notation

$$dA + A \wedge A = F_{ab}dz^{a\alpha} \wedge dz^b{}_\alpha$$

This $F$ can be identified as the self-dual part of the field strength $F_{\mathcal{ABC}}$ defined for the projective light-cone, as follows from identifying the conformal generators in terms of $y$ and $z$, and noting that in spinor notation $F_{ab}$ (symmetric in $ab$) corresponds to a self-dual tensor, while a tensor of the form $F^{ab}$ would correspond to an anti-self-dual one. (There are hidden indices, contracted with group generators. For the case of gravity, these indices carry spin.) The higher-derivative equations of the previous section (covariantizing the derivatives for minimal coupling), which are expressed directly on the on-shell field strengths, are obtained by applying the Bianchi identities to this equation.

The nonzero GL(1) weight of field strengths is indicated by writing them as $F_{(a...b)[\alpha\beta]}$. This is implied by the self-duality equation, but is true for all field strengths, including those without gauge fields (spins 0 and 1/2).

## 7. HARMONICS AND INSTANTONS

These coordinates also are related closely to those used in the harmonic superspace approach to self-dual Yang-Mills theory [13]: There $\lambda^2$ is constrained to 1. As a result, only 7 coordinates (and corresponding gauge connections) are used instead of 8. However, while in the harmonic superspace approach the harmonic coordinates were added by hand in order to increase the light-cone symmetry to the usual Poincaré invariance, from the twistor point of view the appearance of coordinates for SU(2) (the other 3 components of $\lambda$) is automatic, they are part of the same twistor that includes the usual $x$ coordinates, and they are a direct consequence of increasing the manifest symmetry to the conformal group. We now give a brief discussion of this generalization of the harmonic superspace approach, and apply it in a new way to instantons.

The main idea of harmonic superspace is to allow gauge transformations that depend on $\lambda$ as well as $x$. We therefore drop all kinematic constraints on $A$, keeping only those on $F$, which allow this enlarged gauge invariance. In contrast to ordinary gauges, where the covariant derivative satisifes $z^{a\alpha}A_{a\beta} = 0$ to preserve the field independence of the coordinate constraints, in such generalized gauges

$$A = dz^{a\alpha}A_{a\alpha} = dx^{\mu\mu'}\lambda_{\alpha\mu}A_{\mu'}{}^{\alpha} + [(d\lambda^{\alpha}{}_{\nu})\lambda^{-1\nu}{}_{\beta}]z^{a\beta}A_{a\alpha}$$
$$\equiv dx^{\mu\mu'}A_{\mu\mu'} + d\lambda^{\alpha\mu}A_{\alpha\mu}$$

The fact that $A_{\alpha\mu}$ (or $z^{a\beta}A_{a\alpha}$) is pure gauge follows from $z^{a\alpha}F_{ab} = 0$, since this implies the vanishing of field strength terms in the commutation relations of $z^{a\alpha}\nabla_{a\beta}$ with itself and $\nabla$.

As an example of the self-dual twistor approach, we consider the 't Hooft ansatz [14] for SU(2) multi-instanton solutions, which is less general but simpler and more explicit than the ADHM solution. To simplify calculations, it is convenient to generalize the gauge group to SU(2)⊗GL(1) or GL(2), where the extra Abelian piece is pure gauge:

$$A_\iota{}^\kappa = -dz^{a\kappa}\partial_{a\iota}\ln\phi \quad\Rightarrow\quad A_\iota{}^\iota = -d\ln\phi$$

Self-duality then implies

$$F_{ab\iota}{}^\kappa = \tfrac{1}{2}(\partial_{(a}{}^\kappa \ln\phi)(\partial_{b)\iota}\ln\phi)$$

$$\phi^{-1}\partial_a{}^\alpha \partial_{b\alpha}\phi = 0 \quad\Rightarrow\quad \phi = \sum_i (y\cdot y_i)^{-1},\quad y_i^2 = 0$$

(The index $i$ takes k+1 values for k instantons.) We then find

$$z^{a\alpha}\partial_{a\beta}\phi = -\delta^\alpha_\beta \phi \quad\Rightarrow\quad z^{a\alpha}A_{a\beta\iota}{}^\kappa = \delta^\alpha_\iota \delta^\kappa_\beta$$

By comparison with the general expression for $A$ in $\lambda$-dependent gauges, we see that $\lambda$ is itself the gauge parameter that takes $A$ to a gauge where $z^{a\alpha}A'_{a\beta} = 0$:

$$A_\iota{}^\kappa = \lambda^{-1\mu}{}_\iota d\lambda^\kappa{}_\mu + \lambda^{-1\mu}{}_\iota A'_\mu{}^\nu \lambda^\kappa{}_\nu$$

This means that $A$ can be reduced to the usual four-dimensional expression just by setting $\lambda = \delta$. Then

$$A_\iota{}^\kappa = -dx^{\kappa\mu'}\partial_{\iota\mu'}\ln\phi,\quad \phi = -\sum_i 2/e_i(x-x_i)^2$$

The twistor expression is simpler because: (1) the parameters $e_i$ and $x_i$ are unified into the null six-vectors $y_i$, and (2) $\phi$ is homogeneous in $y$ (and in the $y_i$).

This way of writing instantons differs from the usual harmonic superspace methods, where the symmetry on the $\alpha$ indices is explicitly broken down to U(1) or GL(1). In that approach, the gauge $\lambda = \delta$ gives the usual light-cone treatment of self-dual theories. Obtaining a covariant result requires finding the explicit gauge transformation that returns the gauge field to the usual $x$-gauges, a procedure that is difficult in general and has not yet been accomplished for more than one instanton.

One of the most important points about self-dual twistors, which makes them simpler than Minkowski twistors, is that they can be instantly reduced to ordinary four-dimensional coordinates by gauging $\lambda \to \delta$. (This is also true to some extent for self-dual twistors in comparison to projective light-cone coordinates because the latter are constrained, especially in the supersymmetric case.) As a result, many covariant formulas can be reduced by just replacing the four-component spinor indices with two-component ones. The converse is also true, and we will see below how the four-dimensional chiral superspace description of super Yang-Mills can be immediatley generalized to self-dual twistor superspace, where the ADHM construction is simpler.

## 8. ADHM REVIEW

We now review the ADHM construction in a way that emphasizes the conformal symmetry of classical Yang-Mills theory by manifesting it at all stages. Besides slightly simplifying the form of the solution, this clarifies the relationships of the construction to other approaches. (We will not include the proof of completeness.) The first ingredient of the construction is to express the gauge field as that for nonlinear sigma models of coset spaces G/H [15]:

$$A_\iota{}^\kappa = u^I{}_\iota du_I{}^\kappa, \quad u^I{}_\iota u_I{}^\kappa = \delta_\iota^\kappa$$

The indices $\iota, \kappa$ are indices for the defining representation of the Yang-Mills group H. When four-dimensional $x$ space is Euclidean, this group is any of the classical groups USp(2n), SU(n), or SO(n). On the other hand, when $x$ space has two space and two time dimensions, we must use their real Wick rotations Sp(2n) or SL(n) (or still SO(n)). (By USp(2n) we mean unitary symplective 2n×2n matrices, and by Sp(2n) symplective 2n×2n matrices over the reals.) The index $I$ is for the group G: USp(2k+2n), SU(2k+n), or SO(4k+n), etc., where k is the instanton number. (For the USp case we can also use quaternion language, since USp(2m) over the complex numbers is U(m) over the quaternions.) For the SO and (U)Sp cases both kinds of indices can be raised and lowered with the corresponding group metrics (symmetric or antisymmetric, respectively), and $u$ is real (with two real or two pseudoreal indices). For the SU case $u^I{}_\iota$ is the complex conjugate of $u_I{}^\iota$, while for SL they are real and independent (except for the orthonormality relation). We have expressed $A$ as a differential form (or, equivalently, suppressed the index on the derivative $d$) because we will use this same expression for different kinds of coordinates (four-dimensional, projective light-cone, twistor). $u_I{}^\iota$ is thus a part of the matrix representation $U_I{}^{I'}$ ($I' = (\iota, i\alpha)$) of an element of the group G, while $A$ is the part of $U^{-1}dU$ in the Lie algebra of H.

The other ingredient is to complement $u$ with objects $v$ which together form a basis for the space of the $I$ index, and determine the $v$ that describes self-dual solutions. (This is almost the same as choosing $v$ as the missing part of the matrix $U$, but the normalization will be more general to allow simpler coordinate dependence.) The $v$'s are orthogonal to the $u$'s

$$u^I{}_\iota v_{Ii\alpha} = v^I{}_{i'\alpha} u_I{}^\iota = 0$$

and satisfy the same reality properties. The $i, i'$ indices are for the defining representation of GL(k) for the (U)Sp case and GL(2k) for the SO case. For the SU case, $i$ is a GL(k,C) index and $i'$ is the complex conjugate index, while for the SL case $i$ and $i'$ are indices for independent GL(k) groups. The $\alpha$ index is the same as that on $z^{a\alpha}$: In fact, $v$ is required to have the explicit $z$ dependence

$$v_{Ii\alpha} = b_{Iia} z^a{}_\alpha, \quad v^I{}_{i'\alpha} = b^I{}_{i'a} z^a{}_\alpha$$

where the $b$'s are constant and satisfy the same reality properties. Finally, unlike the $u$'s, the $v$'s are not orthonormal, but are required to satisfy a weaker condition:

$$v^I{}_{i'\alpha} v_{Ii\beta} = C_{\beta\alpha} g_{ii'} \quad \Leftrightarrow \quad b^I{}_{i'(a} b_{Iib)} = 0$$

The basic idea of the construction is then to (1) find the most general solution to the constraint on the $b$'s, (2) solve the orthonormality conditions for the $u$'s to express them in terms of the $b$'s, and (3) plug the expression for the $u$'s into the equation $A = \bar{u} du$. The fact that $u$ can be chosen as a function of just $x$ follows from rewriting the $uv$ orthogonality relations as (multiplying by a $z$)

$$u^I{}_\iota b_{Iia} y^{ab} = u_I{}^\iota b^I{}_{i'a} y^{ab} = 0$$

and noting that they are scale invariant. Then also $du$ can be expressed in terms of just $dx$.

We can check the self-duality of the field strength by using the completeness relation that follows from these orthogonality conditions:

$$\delta^J_I = u_I{}^\iota u^J{}_\iota + v_{Ii}{}^\alpha g^{ii'} v^J{}_{i'\alpha}$$

where $g^{ii'}$ is the "metric" inverse to $g_{ii'}$. The field strength is then, in matrix notation,

$$\begin{aligned} F &= dA + A \wedge A \\ &= d\bar{u} \wedge du - d\bar{u}\ u\bar{u} \wedge du \\ &= d\bar{u}\ vg\bar{v} \wedge du \\ &= \bar{u}dv\ g \wedge d\bar{v}\ u \\ &= \bar{u}b\ dz\ g \wedge dz\ \bar{b}u \end{aligned}$$

or more explicitly

$$F_\iota{}^\kappa = g^{ii'}(u^I{}_\iota b_{Iia})(u_J{}^\kappa b^J{}_{i'b})dz^{a\alpha} \wedge dz^b{}_\alpha = F_{ab\iota}{}^\kappa dz^{a\alpha} \wedge dz^b{}_\alpha$$

Thus self-duality follows from the explicit $z$-dependence and quadratic constraint of the $v$'s. Since $u$ is a function of only $x$, and

$$g_{ii'} = \tfrac{1}{2}b^I{}_{i'a}b_{Iib}y^{ab}$$

$F$ is a function of only $x$, up to an overall scale factor (cancelling the same factor in $dz^{a\alpha} \wedge dz^b{}_\alpha$). Furthermore, because of $uv$ orthogonality,

$$z^{a\alpha}F_{ab\iota}{}^\kappa = 0 \quad \Rightarrow \quad y^{ab}F_{bc\iota}{}^\kappa = 0$$

The conformal language is slightly simpler than the usual description because: (1) the parameters $b$ form a single irreducible tensor (or two for SL groups), as does the $\bar{b}b$ constraint, and (2) $v$ is just proportional to $z$. In the usual approach (resulting from the gauge $\lambda = \delta$), breaking up $a = (\mu, \mu')$ divides $b$ and $\bar{b}b$ into separate pieces, while making $v$ linear ($v_\alpha = b_\alpha + b_{\alpha'}x_\alpha{}^{\alpha'}$). This simplification will be magnified in the supersymmetric case, where $z$ breaks up into commuting and anticommuting pieces.

## 9. CONFORMAL CHIRAL SUPERSPACE

The twistor formulation of self-dual Yang-Mills theory makes the supersymmetric generalization obvious. Conversely, the four-dimensional supersymmetric formulation makes the twistor generalization obvious. This is because the reduction from the twistor variables to the usual $x$ results directly from the gauge choice $\lambda = \delta$, whereupon $z^{a\alpha} \to x^{\alpha\mu'}$. For the anticommuting $\theta$ coordinates to be real, supersymmetry requires restriction to the case where $x$ represents two space and two time dimensions. Then all groups, incuding spacetime, internal (SL(N) for N-extended supersymmetry), Yang-Mills, and the dummy groups used in the ADHM construction, are over the reals. (A possible exception is N=2, since SU(2) spinors are pseudoreal.) Since self-dual super Yang-Mills has already been formulated in a chiral superspace with manifest SL(N|2)⊗SL(2) symmetry [16], the reverse of the $z \to x$ reduction means simply extending the super index, to make the symmetry SL(N|4)⊗SL(2), where SL(N|4) is the superconformal group. (In the case N=4, the superconformal group is actually SSL(4|4), where the equality of the bosonic and fermionic range of indices allows an additional tracelessness condition on the generators.)

The four-dimensional chiral superspace formulation used indices $A' = (\mu', a')$ for coordinates $x^{A'\alpha} = (x^{\alpha\mu'}, \theta^{a'\alpha})$ and the corresponding derivatives. The field strengths in this superspace are

$$[\nabla_{A'\alpha}, \nabla_{B'\beta}\} = C_{\beta\alpha} F_{A'B'}$$

where $\nabla_{A'\alpha} = \partial_{A'\alpha} + A_{A'\alpha}$ is expressed in terms of ordinary partial derivatives $\partial_{A'\alpha}$, so there is no torsion term (no $\nabla$ term on the right-hand side). This is a result of having only half the $\theta$'s of the usual superspace. However, no covariance is lost: While only half the supersymmetries (but all the bosonic translations) are described by partial derivatives, the other half are among the fermionic rotations on the $A'$ index. Self-duality is a consequence of having only a $C_{\beta\alpha}$ term in the field strengths (a straightforward generalization of the bosonic case). In ordinary superspace, flat superindices like $A'$ are only formally unified: They are reducible representations of the Lorentz group, not a supergroup, and the bosonic and fermionic parts must be treated separately. They are related only by differential constraints involving torsions. On the other hand, in self-dual chiral superspace there is a graded symmetry for which the super-index labels the defining representation, and there is no torsion to break this symmetry. There the field strength $F_{A'B'}$ is an irreducible representation of this graded symmetry, being just a graded symmetric tensor. Consequently, self-dual chiral superspace is simpler and also manifests more symmetry than the usual superspace.

To generalize to self-dual twistor space, we keep the same equation, but now $z^{A\alpha} = (z^{a\alpha}, \theta^{a'\alpha})$:

$$[\nabla_{A\alpha}, \nabla_{B\beta}\} = C_{\beta\alpha} F_{AB}$$

For the case N=0, this is already the same as the equation used earlier for self-dual Yang-Mills in twistor space. The relation to the four-dimensional coordinates is also obvious: Writing $A = (\mu, A')$,

$$z^A{}_\alpha = \lambda_\alpha{}^\nu (\delta^\mu_\nu, x^{A'}{}_\nu)$$

We can also define a projective super light-cone:

$$y^{AB} = z^{A\alpha} z^B{}_\alpha$$

which is the solution to the constraint [17]

$$y^{[AB} y^{CD)} = 0$$

(where [ ) is graded antisymmetrization and ( ] graded symmetrization). In the gauge $\lambda = \delta$

$$y^{AB} = \begin{pmatrix} C^{\nu\mu} & x^{A'\nu} \\ -x^{B'\mu} & x^{A'\sigma} x^{B'}{}_\sigma \end{pmatrix}$$

The constraints $y^{AB} A_{AB} = 0$, $z^{A\alpha} A_{A\beta} = 0$, etc., are also the direct generalizations of the bosonic twistor expressions obtained by extending the indices $a \to A$. (As usual, our graded Einstein summation convention includes minus signs appropriate to index ordering.)

Thus, by starting with self-dual theories in four-dimensional spinor notation SL(2)⊗SL(2) (for indices on both fields and coordinates), generalizations to both supersymmetry and conformal symmetry result from extending one of the two SL(2)'s to SL(4) (conformal), SL(N|2) (supersymmetric), or SL(N|4) (superconformal). The relationship is not just formal, but is implemented directly on the field equations by simply changing the range of indices. (The conformal generalization also requires additional kinematic constraints.)

## 10. SUPER MULTI-INSTANTONS

The supersymmetric version of ADHM now follows just by extending the indices: Collecting the equations,

$$A_\iota{}^\kappa = u^I{}_\iota du_I{}^\kappa, \quad u^I{}_\iota u_I{}^\kappa = \delta_\iota^\kappa$$

$$u^I{}_\iota v_{I i \alpha} = v^I{}_{i'\alpha} u_I{}^\iota = 0$$

$$v_{I i \alpha} = b_{IiA} z^A{}_\alpha, \quad v^I{}_{i'\alpha} = b^I{}_{i'A} z^A{}_\alpha$$

$$v^I{}_{i'\alpha} v_{I i \beta} = C_{\beta\alpha} g_{ii'} \quad \Leftrightarrow \quad b^I{}_{i'(A} b_{IiB]} = 0$$

$$F_{AB\iota}{}^\kappa = g^{ii'} (u^I{}_\iota b_{IiA})(u_J{}^\kappa b^J{}_{i'B})$$

The main differences now are that (1) the constrained constants $b$ now include both the usual bosonic components and new fermionic components, labeled by $A = a$ and $a'$ respectively, (2) the constraints quadratic in $b$ now include the usual bosonic ones on the bosonic $b$'s as well as both fermionic and bosonic ones on the fermionic $b$'s, labeled by $AB = ab$, $ab'$, and $a'b'$, respectively, and (3) the field strength $F$ not only has extended spacetime indices, but is also a function of $\theta^{a'\alpha}$. The last property means that we obtain expressions for the various physical fields not only by evaluating $F_{AB}$ at $\theta = 0$ for all the nonnegative helicity states, but also by taking covariant $\theta$-derivatives of the scalars $F_{a'b'}$ for the negative helicity states for N>2.

As a consequence of the first two properties, the number of independent bosonic parameters is the same as for the nonsupersymmetric case, but there are also fermionic parameters. As usual, we count degrees of freedom by taking the number of components of $b$ and subtracting the numbers of components of the $\bar{b}b$ constraints and the two dummy symmetry groups:

| | |
|---|---|
| $b$ components | $2(mk+n)k(4\oplus N)$ |
| $\bar{b}b$ constraint | $(mk+m-2)k[5\oplus 2N\oplus \frac{1}{4}N(N-1)]$ |
| GL group(s) | $mk^2$ |
| S group | $(2mk+4n+2-m)k+d(H)$ |
| bosonic parameters | $4(n+2-m)k-d(H)$ |
| fermionic parameters | $2(n+2-m)kN$ |
| less | $\frac{1}{4}(mk+m-2)kN(N-1)$ |

where the number "m" represents the Yang-Mills group $Sp(2n)$ (m=1), $SL(n)$ (m=2), or $SO(n)$ (m=4), while d(H) is the dimension of that group ($n(2n+1)$, $n^2-1$, $\frac{1}{2}n(n-1)$, respectively). The "$4\oplus N$" in $b$ refers to commuting $\oplus$ anticommuting, while the "$5\oplus 2N\oplus\frac{1}{4}N(N-1)$" in $\bar{b}b$ refers to (commuting)$^2$ $\oplus$ commuting $\times$ anticommuting $\oplus$ (anticommuting)$^2$. "Less" is the number of constraints quadratic in anticommuting parameters. These last constraints are difficult to count without explicitly solving the others, and it is possible that they are redundant. (However, there are none for N=1, or for one-instanton $Sp(2n)$.) If so, then the number of fermionic parameters, and the absence of scalar parameters, agrees with the known results [18] for adjoint massless fermions and scalars in background multi-instantons. It also agrees with explicitly supersymmetric one-instanton SL(2) constructions [19], where these parameters can be introduced simply by performing supersymmetry and S-supersymmetry transformations on the purely Yang-Mills result, if we remember to use only chiral supersymmetries (2+2). The explicit expression for the helicity $+1/2$ fields ($F_{ab'}$ at $\theta=0$), and the corresponding N=1 fermionic constraints, agrees with solutions obtained for adjoint fermions in background ADHM instantons [20]. Finally, this super ADHM construction is equivalent to that obtained earlier in nonchiral superspace [21], but the solution here takes a simpler form because: (1) all parameters are combined, as are gauge fields and field strengths, through the use of SL(N|4) indices, and (2) chiral superspace has half the usual superspace coordinates, eliminating the usual expressions quadratic in $\theta$. In particular, in the usual superspace one first solves for the spinor part of the gauge field, and then solves for the vector part as its derivative, while in chiral superspace the whole gauge field comes together automatically.

In self-dual super Yang-Mills theory (and ungauged supergravity), the equation of motion for each helicity involves only that helicity and higher ones, as a consequence of helicity conservation. (This is obvious in the light-cone formalism, where helicity is related to the order in $\theta$, and the field equations have no $\theta$ derivatives.) The self-duality of the Yang-Mills field strength (helicity $+1$) is unaffected by matter, and the helicity $+1/2$ spinor (with multiplicity N) couples only to the Yang-Mills field, while the scalars and negative-helicity fields have also Yukawa and other nonminimal couplings. Explicitly,

$$L = G\tilde{F} + \tilde{\chi}\slashed{\nabla}\chi + \phi\Box\phi + \phi\chi^2$$

$$\Rightarrow \quad \tilde{F} = \slashed{\nabla}\chi = \Box\phi + \chi^2 = \slashed{\nabla}\tilde{\chi} + \phi\chi = \slashed{\nabla}G + \tilde{\chi}\chi + \phi\overleftrightarrow{\nabla}\phi = 0$$

where $\tilde{F}$ is the anti-self-dual part of the Yang-Mills field strength, $\chi$ represents helicity $+1/2$, $\phi$ are the scalars, $\tilde{\chi}$ represents helicity $-1/2$, and $G$ is a Lagrange multiplier representing helicity $-1$. Our results give the solutions for $\phi$ (for N>1), $\tilde{\chi}$ (N>2), and $G$ (N=4) explicitly in terms of the ADHM parameters for $F$ and the analogous parameters for $\chi$.

The 't Hooft ansatz also supersymmetrizes easily in this formalism:

$$A_\iota{}^\kappa = -dz^{A\kappa}\partial_{A\iota}\ln\phi, \quad F_{AB\iota}{}^\kappa = \tfrac{1}{2}(\partial_{(A}{}^\kappa\ln\phi)(\partial_{B]\iota}\ln\phi)$$

$$\phi^{-1}\partial_A{}^\alpha\partial_{B\alpha}\phi = 0 \quad \Rightarrow \quad \phi = \sum_i (\tfrac{1}{2}y^{AB}y_{iAB})^{-1}$$

$$y_{i[AB}y_{iCD)} = 0 \quad \Rightarrow \quad y_{iAB} = \zeta_{iA}{}^{\alpha'}\zeta_{iB\alpha'}$$

in terms of supertwistors $\zeta$ that have chirality opposite to that of $z$. (Summation over $i$ is only where explicitly indicated.) Each twistor parameter appears with its own SL(2) invariance (and $\phi$ itself has a global scale invariance), so there are 5(k+1)−1 bosonic parameters (the usual bosonic result) and 2N(k+1) fermionic ones. The lack of parameters compared to super ADHM is analogous to the bosonic case: For each instanton there are 4 translations, 1 scale, and 2N (chiral) supersymmetry transformation parameters; there are also 4 conformal boosts and 2N S-supersymmetries overall. In contrast, the full solution has for each instanton 4 translations, 1 scale, 3 SL(2), 2N (chiral) supersymmetry, and 2N S-supersymmetry transformations, less 3 overall SL(2) transformations.

Again $A$ can be reduced to a four-dimensional expression just by setting $\lambda = \delta$:

$$A_\iota{}^\kappa = -dx^{A'\kappa}\partial_{A'\iota}\ln\phi$$

Explicitly, we write the dimensional reduction of the coordinates and parameters as

$$z^A{}_\alpha = \lambda_\alpha{}^\nu(\delta^\mu_\nu, x_\nu{}^{\mu'}; \theta^{a'}{}_\nu), \quad \zeta_{iA}{}^{\alpha'} = \mu_i{}^{\alpha'}{}_{\nu'}(-x_{i\mu}{}^{\nu'}, \delta^{\nu'}_{\mu'}; \kappa_{ia'}{}^{\nu'})$$

$$\Rightarrow \quad \phi = -\sum_i 2/e_i (x^{\mu\mu'} - x_i^{\mu\mu'} + \theta^{a'\mu}\kappa_{ia'}{}^{\mu'})^2$$

where $e_i = \mu_i^2$. Twistor notation again unifies all parameters ($e_i$, $x_i$, $\kappa_i$).

## ACKNOWLEDGMENTS

I thank Dileep Jatkar and Vladimir Korepin for discussions. This work was supported in part by the National Science Foundation Grant No. PHY 9309888.